# FACTORS INFLUENCING CITIES' PUBLISHING EFFICIENCY

György Csomós
University of Debrecen, Department of Civil Engineering
csomos@eng.unideb.hu

**Abstract**
Recently, a vast number of scientific publications have been produced in cities in emerging countries. It has long been observed that the publication output of Beijing has exceeded that of any other city in the world, including such leading centres of science as Boston, New York, London, Paris, and Tokyo. Researchers have suggested that, instead of focusing on cities' total publication output, the quality of the output in terms of the number of highly cited papers should be examined. However, in the period from 2014 to 2016, Beijing produced as many highly cited papers as Boston, London, or New York. In this paper, I propose another method to measure cities' publishing performance; I focus on cities' publishing efficiency (i.e., the ratio of highly cited articles to all articles produced in that city). First, I rank 554 cities based on their publishing efficiency, then I reveal some general factors influencing cities' publishing efficiency. The general factors examined in this paper are as follows: the linguistic environment, cities' economic development level, the location of excellent organisations, cities' international collaboration patterns, and the productivity of scientific disciplines.

**Keywords**: Cities, publishing efficiency, spatial scientometrics, highly cited articles, Web of Science

**Introduction**
Both the total publication output of China (Andersson et al. 2014; Grossetti et al. 2014; Morrison 2014; Zhou et al. 2009a) and its publication output in specific research areas (Kumar and Garg 2005; Lu and Wolfram 2010; Zou and Laubichler 2017; Zhou et al. 2009b) have significantly increased in the past decades. The growth rate of China's publication output is quite extreme; however, India (Gupta et al. 2011), Iran (Moin et al. 2005), Brazil (de Almeida and Guimarães 2013; Leta et al. 2006), South Korea (Kim et al. 2012), and Taiwan (Miyairi and Chang 2012) have also recently witnessed significant growth in their total publication output. At the same time, the global share of the publication output of the most developed countries (e.g., the United States, Canada, the Western European countries, Japan, and Australia) has been slowly decreasing. Naturally, the United States still has the highest publication output in the world (Leydesdorff and Wagner 2009; Nature Index 2016), but it can easily be predicted that, due to China's robust growth in the production of science, the global hegemony of the United States will soon cease.

Some cities in the world have long been considered as an outstanding locus of the production of science (Matthiessen and Schwarz 1999; Van Noorden 2010), and for some decades, an increasing number of cities have been involved in that process (Grossetti et al. 2014; Maisonobe et al. 2017). However, cities' contribution to the global publication output has been changing over time. Before the rise of Chinese cities, most global output was primarily produced by Northern American cities (e.g., New York, Boston, and Los Angeles), Western European cities (e.g., London, Paris, and Rome), and Japanese cities (e.g., Tokyo, Kyoto, and Osaka). Currently, Beijing is producing the highest publication output in the world (Csomós 2018; Van Noorden 2010). Furthermore, some cities in emerging countries have been positioning themselves as major actors in the production of science. For example, the publication output of Seoul (South Korea), Tehran (Iran), and São Paulo (Brazil) has also increased significantly.

The question is whether the total publication output clearly represents the scientific performance of a city. Can we find another method to measure the scientific performance of a city, a method that is not based on total (or any kind of) output? Does the geographical pattern of the global production of science change if we focus on quality rather than quantity regarding cities' publication output?

According to Van Noorden (2010), there some alternatives to express the quality of a city's scientific performance, for example, measuring the 'average number of citations that a research paper from a city attracts' or measuring the total number of *Nature* and *Science* articles published by researchers affiliated with that city. Recent studies recommend that, to measure the quality of cities' publication output, the focus should be on the citation impact of the articles published in those cities. According to Bornmann and Leydesdorff (2011), Bornmann and Waltman (2011), Bornmann et al. (2011), and Bornmann and Leydesdorff (2012) as centres of excellence, cities can be assessed by counting the number of excellent papers (i.e., the top 1% most highly cited papers) produced in a city. These studies suggest that, based on the quality of the publication output, cities located in the most developed countries (i.e., the United States, Canada, the Western European countries, Japan, and Australia) are still in top positions.

It is, however, assumed that the higher a city's total publication output is, the more likely it is that the output of highly cited papers will also be high (e.g., currently Beijing produces the greatest number of highly cited papers in the world). This context suggests that, instead of focusing on cities' total publication output or the output of highly cited papers, we should focus on cities' publishing efficiency (i.e., the ratio of highly cited papers to all papers).

Why is it important to measure a city's publishing efficiency? It can be assumed that the higher the ratio of the number of highly cited papers to all articles produced in a city is, the more likely it is that researchers affiliated with that city conduct research resulting in new scientific breakthroughs (Van Noorden's study also suggests this nexus). Thus, publishing efficiency shows how successful a city is at the production of science. In 2015, 2.28 percent of the world's GDP was spent on research and development (R&D) but of course this value varied country to country. In some countries, a higher proportion of the GDP was spent on R&D (e.g., Israel, Japan, and Sweden spent more than three percent of their GDP on R&D), while most countries' R&D expenditures remain under the world average (e.g., the United Kingdom spent less than two percent of their GDP on R&D). Publishing efficiency is a measure that informs governments on how effectively the R&D expenditures have been used (for example, the mean publishing efficiency of UK cities is almost twice as much as that of Japanese cities, while Japan has a much higher R&D expenditure). Furthermore, because publishing efficiency is measured on the city level, it allows governments to introduce more effective regional development policies.

There are many factors influencing cities' publishing efficiency, some of which are city specific and some of which are general. Most of the city-specific factors are related to human factors (for example, how prolific a researcher is), which, due to their nature, vary city to city. However, based on the general factors, typical geographical patterns can be revealed. In this paper, I aim to measure cities' publishing efficiency worldwide and present the most significant general factors that might influence their publishing efficiency.

The structure of the paper is as follows. In Section 2, I present the data collection process and the methodology. Section 3 is divided into two subsections. In the first subsection, I rank cities based on their publishing efficiency, and in the second subsection, the most significant general factors are presented. Finally, in Section 4, I discuss the results and draw the conclusions.

**Data and Methodology**

In the analysis, only cities that had at least 3,000 journal articles published in the period from 2014 to 2016 (i.e., at least 1,000 articles per year) are included. This criterion was met by 554 cities. Data of scientific publications were provided by the Clarivate Analytics' Web of Science database. Two constraints were implemented to improve the objectivity of the study: 1) Only journal articles were selected for the analysis, and 2) journals should be included in the Science Citation Index Expanded (SCI-EXPANDED), the Social Sciences Citation Index (SSCI), and the Arts & Humanities Citation Index (A&HCI) databases.

The reason for the first constraint is that journal articles are generally considered the most prestigious of scientific publications since they are 'the basic means of communicating new scientific knowledge' (Braun et al. 1989: 325). Therefore, I excluded all other types of publications indicated by the Web of Science (e.g., meeting abstracts, book reviews, editorial materials, reviews, proceedings paper, etc.).

The reason for the second constraint is that, in 2015, Clarivate Analytics launched a new database in the Web of Science, the Emerging Sources Citation Index (ESCI), which includes journals of regional importance from emerging scientific fields but that are not yet listed in the Journal Citation Report (i.e., they do not have an impact factor).

The publishing efficiency of a given city (*x*) in the period from 2014 to 2016 (*y*) is obtained by dividing the number of the highly cited articles by the number of all articles produced by authors affiliated with that city (the value is multiplied by 100 to show a percentage). The formula is as follows:

$$Publishing\ Efficiency_{x,y} = \frac{\sum HCA_{x,y}}{\sum A_{x,y}} * 100,$$

where *HCA* is highly cited articles indicated and defined by the Web of Science and *A* denotes all articles indexed by the Web of Science.

The Web of Science presents the name of cities in the addresses reported by the authors of publications. Naturally, one can suggest that it is difficult, if not impossible, to compare scientometric data of cities with very different sizes and populations. For example, Beijing, the Chinese capital, with almost 22 million inhabitants and an area of 16,000 km$^2$ is obviously not on the same tier as Guilford, Surrey, a mid-sized English town with nearly 137,000 inhabitants. The total publication output of Beijing, produced in the period from 2014 to 2016, exceeded that of Guilford by 56 times, and the difference was the same in terms of the number of highly cited articles. However, the publishing efficiency of Beijing and that of Guilford is equal (1.317) since publishing efficiency is calculated as the quotient of the number of highly cited articles and the number of all articles. That is, publishing efficiency is a relative value, and the method it is calculated by makes the size of the city in terms of area or population irrelevant.

It should be noted that the period of data collection spanned from 26/09/2017 to 26/10/2017. The Web of Science has been continuously indexing articles in its database from previous years, especially from 2016. For this reason, in 2017, the number of articles published in 2016 (and before) has been increasing, just like the number of the highly cited articles. A repeated data collection would experience minor differences in the obtained data; however, neither the publishing efficiency nor the rank of cities would considerably change.

**Results**

**Relationship between Cities' Publication Output and Publishing Efficiency**
In the past decades, the publication output of China has radically increased, and the growth rate has exceeded that of any other countries in the world. Furthermore, the publication outputs of some emerging countries, such as South Korea, Taiwan, India, and Iran, have also significantly increased (Csomós 2018; Grossetti et al. 2014; Maisonobe et al. 2017). Naturally, the annual outputs of the United States, Canada, the Western European countries, and Japan have also increased, but they have witnessed a much smaller growth rate than the emerging countries. Therefore, the share of the most developed countries in the production of science has been decreasing for decades (Leydesdorff and Wagner 2009).

Some cities in the world have long been considered as an outstanding locus of the production of science (Matthiessen and Schwarz 1999; Van Noorden 2010), and the growth rate of these cities' publication output is much higher than that of the countries in which they are located. By the beginning of the 2010s, the annual publication output of Beijing surpassed that of any other city in the world (Table 1). In the period from 2014 to 2016, it produced a greater number of scientific publications than Japan. Furthermore, Seoul, Tehran, and São Paulo have also experienced a significant increase in their publication output, and due to this development, their position in the ranking has approached that of Tokyo, Paris, New York, and Boston.

Table 1. Top 50 cities producing the greatest number of articles between 2014 and 2016.

| Rank | Country | City | Total number of articles (2014-2016) | Rank | Country | City | Total number of articles (2014-2016) |
|---|---|---|---|---|---|---|---|
| 1 | China | Beijing | 201260 | 26 | USA | Baltimore, MD | 36528 |
| 2 | China | Shanghai | 98227 | 27 | Germany | Berlin | 36509 |
| 3 | England | London | 92453 | 28 | USA | Philadelphia, PA | 36117 |
| 4 | South Korea | Seoul | 86447 | 29 | China | Chengdu | 36032 |
| 5 | Japan | Tokyo | 77440 | 30 | USA | Houston, TX | 33869 |
| 6 | France | Paris | 75033 | 31 | USA | Atlanta, GA | 32564 |
| 7 | China | Nanjing | 70320 | 32 | Canada | Montreal, PQ | 31820 |
| 8 | USA | New York, NY | 68577 | 33 | China | Tianjin | 31764 |
| 9 | USA | Boston, MA | 63789 | 34 | England | Oxford | 31605 |
| 10 | China | Guangzhou | 51922 | 35 | Germany | Munich | 30886 |
| 11 | China | Wuhan | 50343 | 36 | USA | Seattle, WA | 30779 |
| 12 | Russia | Moscow | 47871 | 37 | Netherlands | Amsterdam | 30498 |
| 13 | Spain | Madrid | 47061 | 38 | USA | Washington, DC | 29986 |
| 14 | Iran | Tehran | 46173 | 39 | Switzerland | Zürich | 29242 |
| 15 | China | Xi'an | 44052 | 40 | Australia | Melbourne, VIC | 29198 |
| 16 | Spain | Barcelona | 40393 | 41 | Sweden | Stockholm | 28599 |
| 17 | Brazil | São Paulo | 39916 | 42 | England | Cambridge | 27907 |
| 18 | USA | Cambridge, MA | 39121 | 43 | China | Changsha | 27442 |
| 19 | China | Hong Kong | 39032 | 44 | USA | Ann Arbor, MI | 27322 |
| 20 | China | Hangzhou | 39029 | 45 | Japan | Osaka | 26594 |
| 21 | USA | Los Angeles, CA | 38740 | 46 | China | Jinan | 26557 |
| 22 | Canada | Toronto, ON | 38497 | 47 | China | Harbin | 26386 |
| 23 | Australia | Sydney, NSW | 37676 | 48 | Denmark | Copenhagen | 25538 |
| 24 | USA | Chicago, IL | 37560 | 49 | Italy | Rome | 25378 |
| 25 | Singapore | Singapore | 37523 | 50 | China | Hefei | 24911 |

However, many researchers have doubts about the quality of publications produced in Brazilian, Chinese, Indian, Iranian, and even South Korean cities, which is also reflected in their low citation impact (Andersson et al. 2014; Maisonobe et al. 2017; Xie et al. 2014; Van Noorden 2010; Zhou et al. 2009a). In the period from 2014 to 2016, the greatest number of highly cited articles were produced in Beijing (see Table 2), which is not surprising, if we consider the extremely high total publication output of Beijing in terms of the number of articles. Table 2 shows that the difference between the output of Beijing and Boston in terms of the number of highly cited articles is very small, while the total output of Beijing is three times greater than that of Boston (see Table 1). Comparing the rankings in Tables 1 and 2, the positions of some top-ranked cities in terms of total output (e.g., Tokyo and Seoul) have dropped in the ranking of cities producing the greatest number of highly cited articles. In addition, such emerging cities, such as Tehran and São Paulo, which both produced a substantial number of articles between 2014 and 2016, have disappeared from the ranking of the top 50 cities with the greatest number of highly cited articles.

Table 2. Top 50 cities producing the greatest number of highly cities articles between 2014 and 2016.

| Rank | Country | City | Number of highly cited articles (2014-2016) | Rank | Country | City | Number of highly cited articles (2014-2016) |
|---|---|---|---|---|---|---|---|
| 1 | China | Beijing | 2650 | 26 | Spain | Madrid | 772 |
| 2 | USA | Boston, MA | 2387 | 27 | USA | Ann Arbor, MI | 765 |
| 3 | England | London | 2337 | 28 | Germany | Munich | 764 |
| 4 | USA | New York, NY | 2237 | 29 | Japan | Tokyo | 734 |
| 5 | USA | Cambridge, MA | 1827 | 30 | Switzerland | Zürich | 730 |
| 6 | France | Paris | 1601 | 31 | Denmark | Copenhagen | 729 |

| 7 | China | Shanghai | 1208 | 32 | Australia | Sydney, NSW | 723 |
| 8 | USA | Seattle, WA | 1191 | 33 | China | Hong Kong | 720 |
| 9 | USA | Los Angeles, CA | 1142 | 34 | Sweden | Stockholm | 708 |
| 10 | England | Oxford | 1083 | 35 | South Korea | Seoul | 698 |
| 11 | USA | Stanford, CA | 1058 | 36 | Germany | Berlin | 678 |
| 12 | USA | Philadelphia, PA | 1050 | 37 | USA | Washington, DC | 674 |
| 13 | USA | Baltimore, MD | 1047 | 38 | USA | Durham, NC | 660 |
| 14 | Canada | Toronto, ON | 1024 | 39 | USA | New Haven, CT | 659 |
| 15 | USA | Chicago, IL | 991 | 40 | Canada | Montreal, PQ | 656 |
| 16 | USA | Atlanta, GA | 971 | 41 | Australia | Melbourne, VIC | 643 |
| 17 | USA | Houston, TX | 964 | 42 | China | Wuhan | 637 |
| 18 | Spain | Barcelona | 920 | 43 | Germany | Heidelberg | 627 |
| 19 | USA | Berkeley, CA | 911 | 44 | Canada | Vancouver, BC | 615 |
| 20 | USA | San Francisco, CA | 910 | 45 | USA | Pittsburgh, PA | 588 |
| 21 | England | Cambridge | 885 | 46 | USA | Princeton, NJ | 587 |
| 22 | Singapore | Singapore | 871 | 47 | China | Guangzhou | 552 |
| 23 | China | Nanjing | 866 | 48 | Switzerland | Geneva | 534 |
| 24 | USA | Bethesda, MD | 825 | 49 | Saudi Arabia | Jeddah | 532 |
| 25 | Netherlands | Amsterdam | 806 | 50 | USA | St. Louis, MO | 511 |

A different geographical pattern will emerge if we focus on measuring cities' publishing efficiency (Figure 1). The mean publishing efficiency of the 554 cities included in the analysis is 1.818, which means that an average of 1.818% of all articles published in these cities in the period from 2014 to 2016 received enough citations to belong to the top 1% of highly cited articles. However, there are significant geographic differences behind the mean value. Figure 1 shows that the publishing efficiency of most Chinese, Japanese, and South Korean cities (many of which have high publication output in terms of the number of articles) is quite low, while the publishing efficiency of most Northern American and Western European cities is considerably higher. This information is not novel since, directly or indirectly, it has also been described by Van Noorden (2010), Bornmann and Waltman (2011), and Leydesdorff et al. (2014).

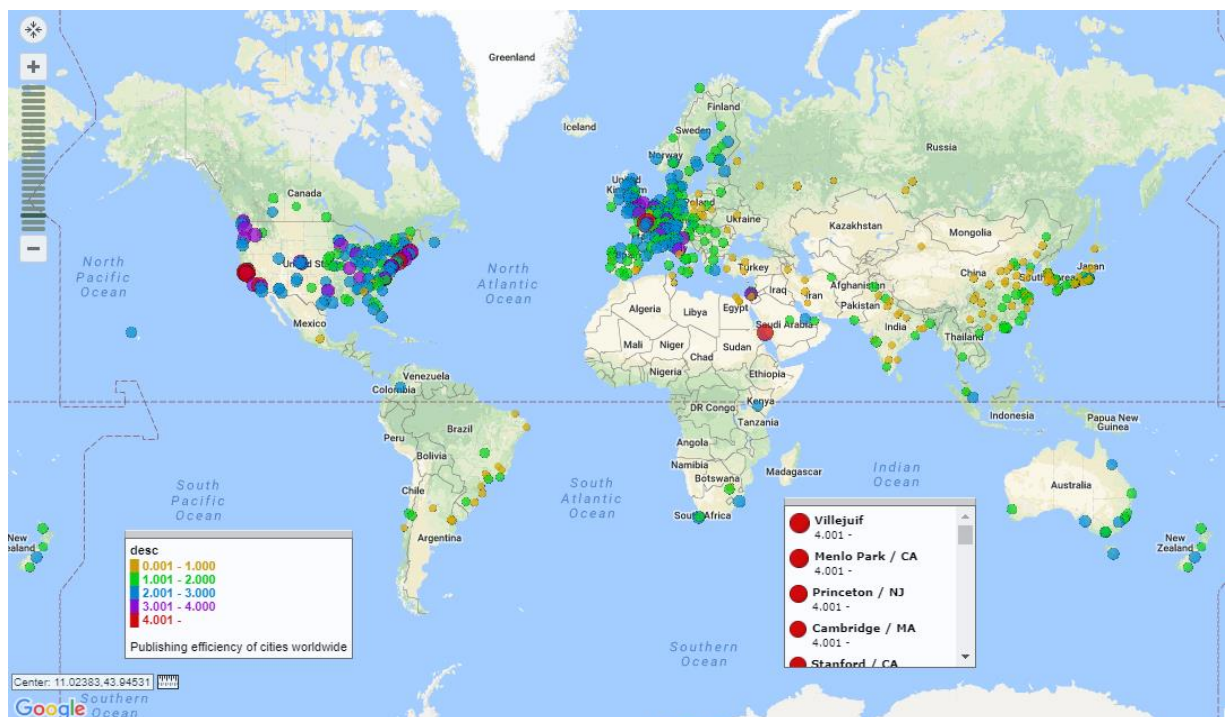

Fig. 1. Geographic visualisation of cities' publishing efficiency.

However, a more fundamental question is whether there are factors influencing cities' publishing efficiency. Are there any general factors producing high publishing efficiency? Can we find general factors characterising cities having low publishing efficiency? Why is the publishing efficiency of Villejuif (France), Menlo Park, California (United States), or Jeddah (Saudi Arabia) high, and what are the reasons behind the low publishing efficiency of Tehran (Iran), Shenyang (China), and Niigata (Japan)? To answer these questions, we should explore and compare the general factors characterising the most efficient and least efficient cities.

The mean publishing efficiency of the top 100 most efficient cities is 3.179, while that of the bottom 100 least efficient cities is 0.621. In the top 100 cities, in the period from 2014 to 2016, a mean of 13,830 articles per city was produced, of which a mean of 444.71 articles per city received enough citations to belong to the top 1% of highly cited articles. In the same period, in the bottom 100 cities, a mean of 8,885 articles per city was produced, of which only a mean of 60.44 articles per city received enough citations to belong to the top 1% of highly cited articles. That is, the total output in terms of the number of articles of the top 100 most efficient cities is only 1.5 times greater than that of the bottom 100 least efficient cities. In contrast to the results above, there is a difference of more than 7.4 times between the number of highly cited articles produced in the top 100 cities and those produced in the bottom 100 cities. When exploring the general factors influencing cities' publishing efficiency, I will focus on presenting the differences between the top 100 most efficient cities and the bottom 100 least efficient cities.

The general factors examined in this paper are as follows: the dominance of the English language and cities' economic development level (both derived from country-level data), the location of excellent organisations, cities' international collaboration patterns, and the productivity of specific research areas. The full list of the top 100 most efficient cities is available in Appendix 1, and the list of the bottom 100 least efficient cities is in Appendix 2.

**Exploring Factors Influencing the Cities' Publishing Efficiency**

Before exploring and evaluating the general factors influencing cities' publishing efficiency, it is necessary to present the geographical location of cities included in the analysis. The geographical location of a given city does not directly influence its publishing efficiency but allows us to draw indirect conclusions.

Most cities producing high publication output in terms of the number of articles (i.e., at least 3,000 articles in the period from 2014 to 2016) are in three geographical regions in the world: Europe, Asia, and Northern America (Table 3). The aggregate proportion of cities from other regions i.e., Africa, Latin America, and Australia/New Zealand) does not reach 9%. Not just the output but also the mean publishing efficiency of cities differs from each other depending on where they are located. Northern American cities produce the highest publishing efficiency, which is almost one-third greater than that of the European cities ranked second. However, if we divide Europe, the most complex region (there are 29 European countries in the analysis), into sub-regions, we obtain a more realistic picture. The mean publishing efficiencies of the Northern European and the Western European cities are much higher than that of the Southern European and Eastern European cities, and while the publishing efficiencies of the former groups approach the efficiencies of the Northern American cities, those of the Eastern European cities are rather close to the efficiencies of the Latin American cities. In Asia, significant differences emerge as well. The mean publishing efficiencies of cities in Southern Asia and Eastern Asia are under the mean efficiencies of Western Asian cities. Furthermore, cities located in the former two Asian sub-regions produce the lowest mean publishing efficiencies in the world.

Table 3. Number of cities and their mean publishing efficiencies by region and sub-region*.

| Regions/Sub-regions | Number of cities | Percentage in the dataset | Cities' mean publishing efficiency |
|---|---|---|---|
| Africa | 11 | 1.99 | 1.306 |
| Asia | 131 | 23.65 | 1.009 |
| *Eastern Asia* | *88* | *15.88* | *0.950* |
| *Southern Asia* | *24* | *4.33* | *0.876* |
| *Western Asia* | *14* | *2.53* | *1.521* |
| Europe | 230 | 41.52 | 1.948 |
| *Eastern Europe* | *25* | *4.51* | *0.989* |
| *Northern Europe* | *60* | *10.83* | *2.260* |
| *Southern Europe* | *54* | *9.75* | *1.673* |
| *Western Europe* | *91* | *16.43* | *2.168* |
| Latin America | 21 | 3.79 | 0.952 |
| Northern America | 145 | 26.17 | 2.497 |
| *Canada* | *18* | *3.25* | *1.970* |
| *USA* | *127* | *22.92* | *2.572* |
| Australia/New Zealand | 16 | 2.89 | 1.918 |
| **World** | **554** | | **1.818** |

*Regions and sub-regions are defined by the United Nations Statistics Division in its geoscheme.

Figure 2 shows the geographical location of the top 100 most efficient cities. Most of the top 100 cities are in two major regions: Northern America (primarily in the United States) and Europe (primarily in Northern Europe and Western Europe). In this group, only three cities are outside the above regions: two of them are in Southern Africa (more precisely in South Africa), and two of them can be found in Western Asia (Saudi Arabia and Israel).

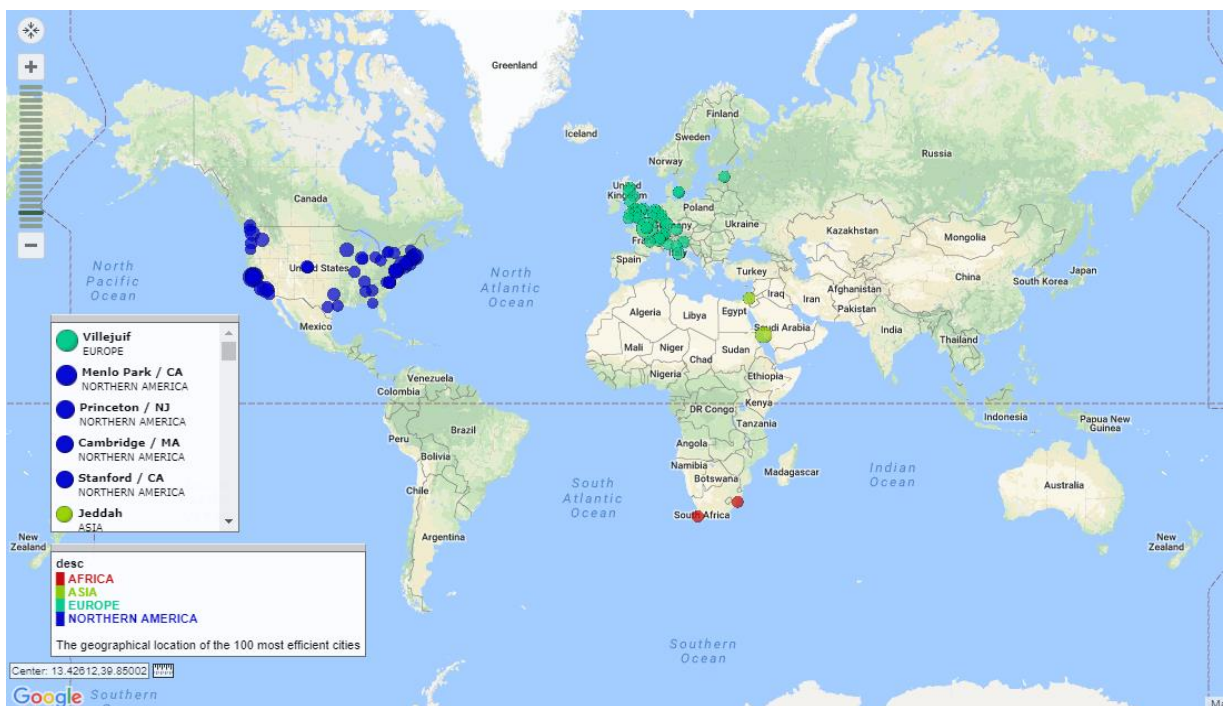

Fig. 2. Geographical location of the top 100 most efficient cities.

The bottom 100 least efficient cities are primarily in three major regions in the world: Asia (primarily in Eastern Asia and Southern Asia), Europe (primarily in Eastern Europe), and Latin America. There is no Northern American city among the least efficient cities, and only two cities from Northern Europe and Western Europe belong to this group. Compared to the number of cities from other regions in the world, the number of African cities (6 cities) is insignificant in this group; however, 55% of the African cities belong to the bottom 100 least efficient cities.

The geographical location of the top 100 most efficient cities and the bottom 100 least efficient cities is indicative information but allows us to deduce some of the general factors influencing cities' publishing efficiency. One of the most crucial factors is related to linguistic features, more precisely to the dominance of the English language.

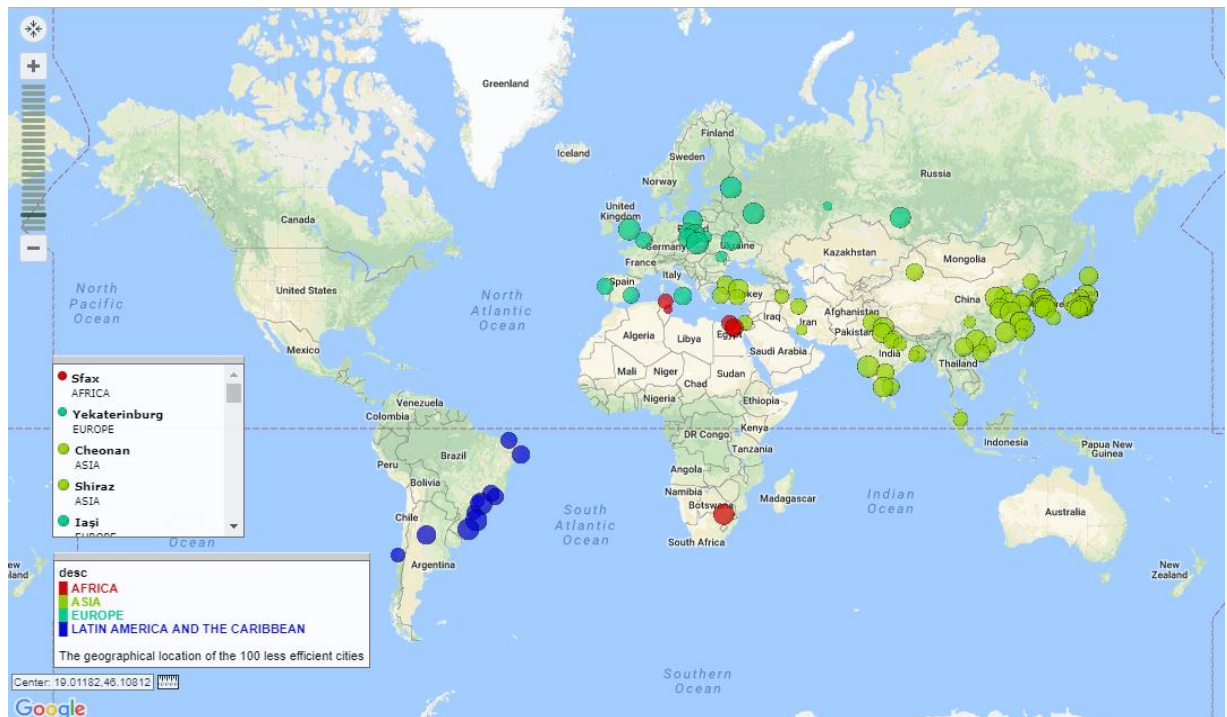
Fig. 3. Geographical location of the bottom 100 least efficient cities.

**The Linguistic Environment as a Factor Influencing Cities' Publishing Efficiency**
It is a generally accepted fact that English has acquired almost exclusive status as the international language of scientific communication (i.e., the neutral "lingua franca"), leaving little space for other languages in science (Björkman 2011; López-Navarro et al. 2015; Tardy 2004; van Weijen 2012). Although the most important indexing and abstracting databases (i.e., the Web of Science and Scopus) have been including an increasing number of non-English language journals, English language journals are still significantly overrepresented (Li et al. 2014; Mongeon and Paul-Hus 2016). According to Paasi (2005), 'Anglo-American journals dominate the publishing space in science', and the international journal publication space is 'particularly limited to the English-speaking countries'. Furthermore, as Braun and Diósptonyi (2005), Braun et al. (2007), and Leydesdorff and Wagner (2009) asserted in terms of gatekeepers like editors-in-chief and editorial board member positions, the dominance of the United States is still unchallenged. Considering the above facts, the English language is assumed to be one of the most principal factors that influence cities' publishing efficiency.

In this paper, I classified cities according to the Anglosphere system introduced by Bennet (2007: 81-83). In this system, the United States, the United Kingdom, Canada, Australia, New Zealand, and Ireland belong to the Anglosphere−Core. Countries in the Anglosphere−Middle sphere (e.g., Nigeria and South Africa) have several official languages, including English (which is the principal language of administration and commerce), but 'where the primary connections to the outside world are in English'. The Anglosphere−Outer sphere consists of English-using states of other civilisations, including India, Pakistan, Bangladesh, the Arab states formerly under British control (primarily in the Middle East), and the Islamic former colonies of Britain (e.g., Malaysia and African states).

A total of 230 cities (out of 554 cities included in the analysis) are in countries in the Anglosphere, from which 195 cities are in countries in the Anglosphere−Core. Countries outside the Anglosphere are home to 324 cities. The mean publishing efficiency of cities in countries in the Anglosphere is 2.271, while that of the rest of the cities is 1.497. That is, the mean publishing efficiency of cities in countries in the Anglosphere is greater than that of the rest of the cities by 50%. If we focus on the mean publishing efficiency of cities located in countries in the Anglosphere−Core, it increases to 2.439.

As for the top 100 most efficient cities, 73% of them are in countries in the Anglosphere, and 70% of them can be found in countries in the Anglosphere–Core. The mean publishing efficiency of cities belonging to the latter group is 3.235. In contrast, 85% of the bottom 100 least efficient cities are in countries outside the Anglosphere, and 99% of them are in countries outside the Anglosphere−Core. Loughborough (England), having a publishing efficiency of 0.868, is the only city in the group of the bottom 100 cities that can be found in the Anglosphere–Core.

In conclusion, the publishing efficiency of cities located in countries in the Anglosphere (especially in the Anglosphere−Core) is much higher than that of any other cities located in countries outside the Anglosphere. That is, English is not only the international language of scientific communication but also the most fundamental factor influencing cities' publishing efficiency.

**Economic Development Level of Cities as a Factor Influencing Publishing Efficiency**
Some researchers have observed linear correlation between scientometric indicators (e.g., the number of publications) and economic development indicators (e.g., GDP per capita or income per capita) (de Solla Price 1978; Kealey 1996; King 2004), while others believe that the correlation between these different sets of indicators is far from clear (Lee at al. 2011; Meo et al. 2013; Vinkler 2008; Vinkler 2010). It is, however, more commonly accepted that the higher the GDP per capita or the income level of a country is, the more likely it is that a greater number of publications will be produced in that country. The question is whether there is a correlation between cities' publishing efficiency (as a scientometric indicator) and cities' per capita income level (derived from country-level data).

The classification of countries (and cities) by income level is based upon data obtained from the World Bank Country and Lending Groups database. In this database, countries are classified into four income-level groups: low-income countries (GNI per capita of $1,005 or less in 2016), lower middle-income countries (GNI per capita between $1,006 and $3,955), upper middle-income countries (GNI per capita between $3,956 and $12,235), and high-income countries (GNI per capita of $12,236 or more).

Results show that 434 out of 554 cities included in the analysis are in high-income countries, 93 of them are in upper middle-income countries, and only 27 cities can be found in lower middle-income countries. None of the cities are in low-income countries. That is, most cities producing high publication output in terms of the number of articles (i.e., at least 3,000 articles in the period from 2014 to 2016) are in high-income countries. The mean publishing efficiency of cities from high-income countries is 2.057, that of cities located in upper middle-income countries is 0.997, and the mean publishing efficiency of cities from lower middle-income countries is only 0.881. There is a difference of more than double between the mean publishing efficiency of cities located in high-income countries and that of cities located in upper middle-income countries. The difference between the mean publishing efficiency of cities in upper middle-income countries and that of cities in lower middle-income countries seems to be insignificant.

As for the top 100 most efficient cities, 98% of them are in high-income countries, and only 2% of them can be found in upper middle-income countries. As compared to the quasi-homogeneous group of the top 100 cities, the bottom 100 least efficient cities show a very complex picture; 18% of them are in lower middle-income countries, and 46% of them are in upper middle-income countries, but 36% of the bottom 100 least efficient cities are in high-income countries. Based on former studies available in the literature, this latter result might not have been expected; therefore, it requires more explanation.

As was mentioned above, most of the top 100 cities were in Northern America (primarily in the United States) and Europe (primarily in Northern European and Western European countries). Almost all countries in these regions are high-income countries. Contrary to the most efficient cities, none of the least efficient cities are in Northern America. Furthermore, only 17% of the bottom 100 cities are in European countries; except for five cities, all of them are in Eastern European countries (including Russia). Results show that 11 out of the 17 least efficient European cities are in high-income countries, and six of them are in Poland. Figure 4 illustrates that many cities producing low publishing efficiency are in Eastern Asian high-income countries. Half of these cities are in South Korea (11 cities), and another half are in Japan (12 cities); i.e., in countries that belong to the most developed countries in the world in terms of income level. One might suggest that if South Korea and Japan are high-income countries, cities located in South Korea and Japan should produce high publishing efficiency. However, both the Korean and Japanese languages are considered language isolates. According to Campbell (2010: 16), a language isolate is a language that has no relatives, that is, no demonstrable genetic relationship with any other language. The Korean language is a language isolate, and Japanese would also be a language isolate if Ryukyuan languages had not been shown to be distinct languages, related to Japanese. However, from the perspective of scientific communication, Japanese is considered a 'quasi' language isolate. This study shows that the linguistic environment where a city is located in plays a more significant role in publishing efficiency than the economic development level of countries in which the cities are located. Loughborough (England) is the only city in the bottom 100 least efficient cities that is in a high-income country belonging to the Anglosphere−Core. Beer-Sheva (Israel), a city in the group of the bottom 100 least efficient cities, is also in a high-income country, but is in the Anglosphere−Outer sphere. In fact, many of the bottom 100 cities are in countries in the Anglosphere−Outer sphere, but all of them are in lower middle-income countries, primarily in Southeast Asia (11 cities are in India, and one is in Pakistan).

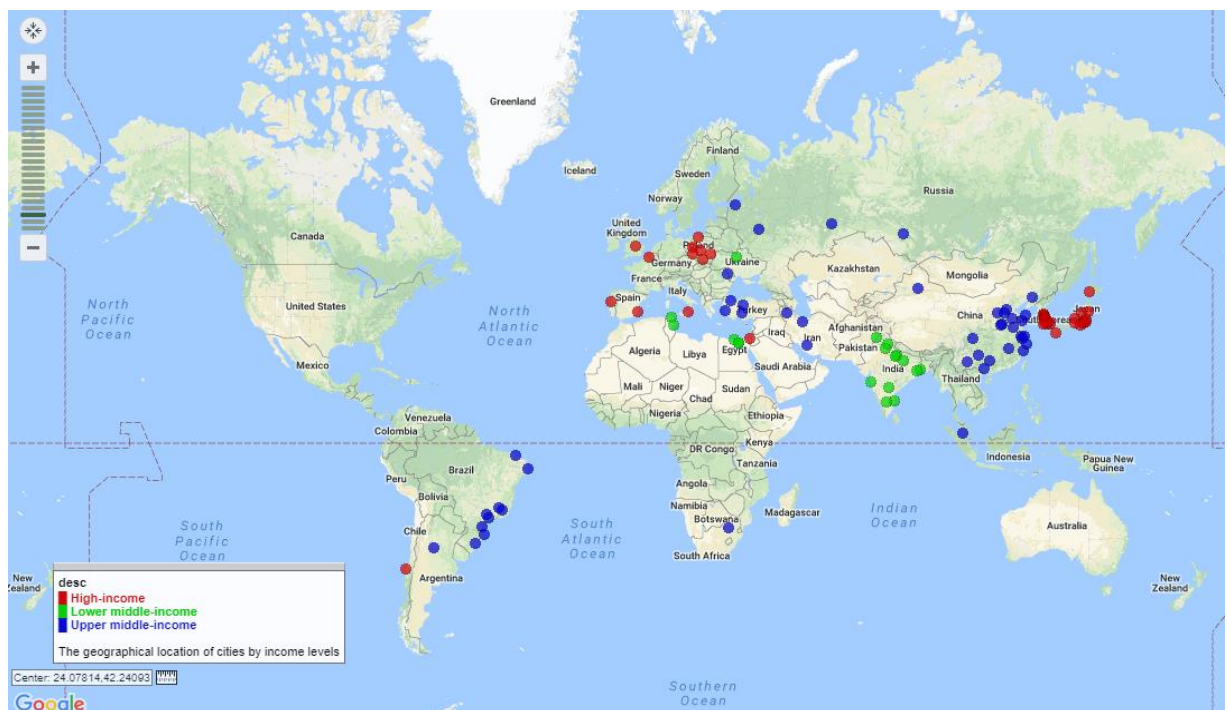

Fig. 4. Geographical location of the bottom 100 least efficient cities in terms of countries' income levels.

East Asia is home to 46% of the bottom 100 least efficient cities. Beside Japan and South Korea, most of these cities are in China. While none of the East Asian countries included in the analysis belong to the group of the low-income or lower middle-income countries (Japan and South Korea are high-income countries, and China is an upper middle-income country), the publishing efficiency of the East Asian cities is rather low. Kawaguchi (Japan), the city producing the highest publishing efficiency in the region, is ranked only 138$^{th}$. The facts above suggest that the economic development level of the cities is a key factor influencing publishing efficiency, which is reinforced by the fact that almost all cities in the group of the top 100 cities are in high-income countries, but it is not the most important factor.

    The examination of factors like the dominance of the English language and cities' economic development level will bring us closer to understanding why cities' publishing efficiency differs from each other; however, we need deeper insight to obtain a precise picture of publishing efficiency. For example, country-level data allows us to understand why the publishing efficiency of Canadian and

Chinese cities significantly differ from each other but does not help us to understand why the publishing efficiency of Kawaguchi is higher or why that of Niigata is lower than the mean publishing efficiency of Japanese cities. To examine cities' publishing efficiency in a more precise way, we need to focus on some general as well as more city-specific factors, like the location of excellent organisations, cities' international collaboration patterns, and the productivity of specific research areas.

For example, in Kawaguchi, most publications were produced by the Japan Science and Technology Agency, one of Japan's excellent scientific organisations; therefore, the publishing efficiency of Kawaguchi is considerably higher than that of other Japanese cities. That is, which cities in the world are home to excellent organisations (e.g., universities and governmental and international research institutions) should be examined. The question is whether these organisations are exclusively located in cities producing high publishing efficiency or whether some of them might be found in cities with low publishing efficiency.

**Location of Excellent Organisations as a Factor Influencing Cities' Publishing Efficiency**
In the paper by Van Noorden (2010: 907) an important question arose: What is the reason Boston ranks top in several analyses of scientific quality? A brief answer was given by José Lobo, a statistician and economist who was affiliated with Arizona State University at Tempe: 'Take three or four of the best universities in the world, put them in a city with a seaport, and voilà!' Naturally, the question requires a more complex answer (as was later also explained by Van Noorden), but it calls attention to a key factor: the scientific performance of cities significantly depends on whether they are home to excellent universities.

Although many research institutions, hospitals, governmental organisations (e.g., ministries and departments), NGOs, and companies have a significant publication output (see, for example, Archambault and Larivière 2011; Csomós and Tóth 2016; Hicks 1995), scientific publications are primarily produced by universities all over the world. In recent years, university rankings have gained in popularity. The main goal of ranking and comparing universities in terms of scientific output (of which the publication output is a vital component) is to make the most excellent universities visible worldwide. There are several different world university rankings available (e.g., CWTS Leiden Ranking, The Times Higher Education World University Rankings, QS World University Rankings, and Academic Ranking of World Universities – ARWU), which are all based upon different input data. However, each ranking attributes more or less significance to bibliometric indicators, such as the number of publications produced in a given university, the quality (citation impact) of scientific publications, or the number of articles published in top journals (see, for example, Docampo et al. 2015; Frenken et al. 2017; Piro and Sivertsen 2016; Shehatta and Mahmood 2016). Naturally, the methodologies of how university rankings are produced differ from each other; thus, university rankings are different in terms of top university rankings (Abramo and D'Angelo 2015; Lin et al. 2013).

From the point of view of this analysis, university rankings contain indicative information only; therefore, I chose to use the Academic Ranking of World Universities (ARWU) published annually by the Shanghai Ranking Consultancy. I examined whether there is a correlation between the location of excellent universities and cities' publishing efficiency. Excellent universities correspond to universities having been ranked among the top 100 universities on one of the ARWU lists of 2014, 2015, and 2016.

In the period from 2014 to 2016, the top 100 universities were in 95 cities, some of which were home to more than one excellent university (e.g., New York, London, Boston, Pittsburgh, Munich, Stockholm, and Zurich). The publishing efficiency of cities that were home to the top 100 universities averages 2.641, while that of the rest of the cities averages 1.648. That is, the mean publishing efficiency of cities that are home to the top 100 universities is higher than that of the rest of the cities by 60%. These results suggest that the location of excellent universities significantly influences cities' publishing efficiency. In other words, it seems to be a logical assumption that excellent universities are primarily located in the most efficient cities. Thus, we should examine which of the top 100 most efficient cities are home to excellent universities.

Figure 5 shows that it is not an exclusive privilege of the most efficient cities to be home to excellent universities. Only 43% of the top 100 universities are in the top 100 most efficient cities. Furthermore, there are many cities worldwide (including Chinese and Japanese cities), that do not belong to the top 100 most efficient cities; yet, they are home to excellent universities. In the group of the bottom 100 cities, Moscow (Russia) is the only city that is home to an excellent university.

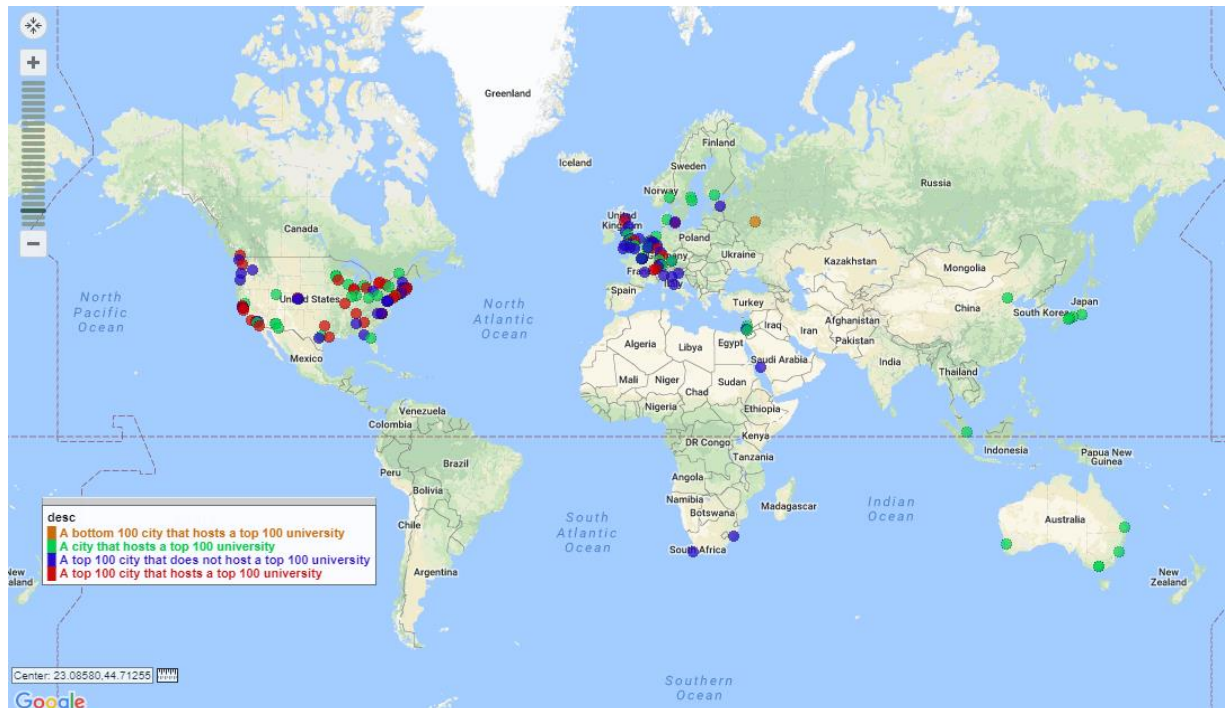

Fig. 5. Geographical location of cities that are home to the top 100 universities as ranked by ARWU.

The location of excellent universities is considered an important but not decisive factor influencing cities' publishing efficiency. Examining the ranking of the most efficient cities, there are two cities (Villejuif, France and Menlo Park, California, USA) topping the ranking that are not home to excellent universities as ranked by the ARWU. It should be noted however, that ARWU is just one of the alternatives to rank universities. Naturally, other organisations produce different rankings with different universities in top positions. For example, out of the top 10 universities, only Harvard University and Stanford University appear in both the CWTS Leiden Ranking of 2017 and the ARWU list of 2017. Contrary to this example, the groups of the top 10 universities in the QS World University Rankings of 2017 and the ARWU list differ from each other by only three universities. In addition, there are many excellent universities that are not included in the group of top 100 universities on the ARWU list but are in cities with high publishing efficiency. For example, Rotterdam, the forty-second most efficient city in the world, is home to the Erasmus University Rotterdam, which ranked 101-150 (i.e., outside but close to the top 100 universities).

The question arises as to what kind of organisations (but not universities) are in cities like Villejuif, Menlo Park, California, Upton, New York (United States), Greenbelt, Maryland (United States), Didcot (England), etc., which produce very high publishing efficiency. The explanations are as follows.

Villejuif, the city with the highest publishing efficiency in the world, is home to the 'Institut Gustave Roussy', one of the world's leading cancer-research institutions and the premier oncology centre and teaching hospital in Europe. Although Villejuif is a city (commune) having 50 thousand inhabitants, it is a suburb of Paris, about seven kilometres from its centre.

Menlo Park is home to the SLAC National Accelerator Laboratory, a linear accelerator that is owned by the US Department of Energy and operated by the Stanford University. Currently, SLAC is the world's largest linear accelerator and is one of top research centres for accelerator physics. The city of Menlo Park, with a population of 32 thousand, is in the San Francisco Bay Area between San Francisco and San Jose (i.e., in one of the fastest growing regions in the world that is home to many innovative companies and top universities). Additionally, Didcot has 25 thousand inhabitants and is 16 km south of Oxford. Didcot is home to the Rutherford Appleton Laboratory, a world-renowned research centre for particle physics and space science.

Cities such as Villejuif, Menlo Park, and Didcot can be characterised the same way; they are smaller cities, towns, or villages located in metropolitan areas and are home to quasi-independent research institutions (e.g., national laboratories) generally operating under the umbrella of excellent universities. Naturally, top research institutions are in large cities as well, but being surrounded by universities, their visibility in terms of publication output is much lower, even if they produce very high publishing efficiency. For example, the total publication output of Geneva (Switzerland) is produced by many organisations, including the European Organisation for Nuclear Research (CERN), the World Health Organization (WHO), and the University of Geneva. In the period from 2014 to 2016, almost 60% of Geneva's total publication output came from the University of Geneva, which has been ranked among the top 100 universities on the ARWU list, and which publishing efficiency is as high as 3.33. However, if we compare the publishing efficiency of the University of Geneva to that of the CERN (5.37) and the WHO (6.86), it seems rather low. The same pattern appears in large cities like New York, London, Paris, Los Angeles, and Tokyo.

In conclusion, a positive correlation can be detected between the location of excellent universities and cities' high publishing efficiency. However, it should be noted that publications, primarily in large cities, come from different types of organisations, many of which have lower publishing efficiency than universities. Thus, some cities that are home to excellent universities have not been included in the top 100 most efficient cities. Furthermore, there are several top-ranked cities that are not home to excellent universities (or any universities); yet, they produce a very high publishing efficiency.

**International Collaboration Pattern as a Factor Influencing Cities' Publishing Efficiency**
In recent years, the number of publications produced by single authors has been decreasing, while the number of co-authored publications and number of co-authors in publications have been increasing rapidly (Abramo et al. 2017; Castelvecchi 2015; Uddin et al. 2012). Therefore, cities' international collaboration patterns have become more complex (i.e., authors affiliated with a given city have been collaborating with a growing number of co-authors affiliated with other cities in other countries). Naturally, cities' international collaboration patterns are influenced by many factors, including differences between the productivity of scientific disciplines (Larivière et al. 2006; Paul-Hus et al. 2017; Zhou et al. 2009b), the size of the national research system (Van Raan 1998), and linguistic features (Csomós 2018; Maisonobe et al. 2016). These facts might suggest that international collaboration patterns vary city to city worldwide, making it impossible to predict cities' publishing efficiency. However, this question remains to be answered.

In this section, I aim to examine whether cities with high publishing efficiency and cities with low publishing efficiency are characterised by general international collaboration patterns. Data obtained from the Web of Science database allows us to reveal countries with which the co-authors are affiliated. For example, in the period from 2014 to 2016, 27,322 articles were produced in Ann Arbor, Michigan (United States), from which 765 received enough citations to belong to the top 1% highly cited articles. If we focus on the international collaboration pattern of all articles produced in Ann Arbor between 2014 and 2016, 8.76% of the articles were written with co-authors affiliated with China, 7.23% had co-authors affiliated with Canada, 7.13% had co-authors affiliated with England, 6.78% had co-authors affiliated with Germany, 5.16% had co-authors affiliated with France, and so on. That is, in the case of all articles, the top collaborator with Ann Arbor is China, and the second ranked collaborator is Canada, and so on.

However, if we focus on the international collaboration pattern of the highly cited articles, a different pattern will emerge. Most highly cited articles were written with co-authors affiliated with England (27.32%), with 25.49% from Canada, 23.53% from Germany, 20.26% from France, 17.39% from Italy, and so on. That is, in the case of highly cited articles, the top collaborator of Ann Arbor is England (replacing China as the top collaborator in all articles), and the second ranked collaborator is Canada, and so on.

I examine which countries are the top collaborators (i.e., collaborators ranked 1-5) in the case of all articles and in the case of highly cited articles produced in a given city in the period from 2014 to 2016. Furthermore, I compare the typical international collaboration patterns of the top 100 most efficient cities to that of the bottom 100 least efficient cities. My aim is to reveal whether there is a correlation between cities' international collaboration patterns and cities' publishing efficiencies and whether there is a difference between the typical collaboration patterns of the top 100 cities and the bottom 100 cities. When examining cities' international collaboration patterns, I implemented a geographical constraint. The group of the top 100 cities was divided into two sub-groups (i.e., the most efficient non-US cities and the most efficient US cities), and they were examined separately.

Table 4 shows the countries occupying the top 1-5 positions as collaborators in all articles and their frequency of occurrence in those positions. The top collaborator of the most efficient non-US cities (48 out of the top 100 cities) is the United States, whose frequency of occurrence in the top 1-5 positions is 100% (in the top position in 81.25% of the cases). This means that the United States has a very intense collaboration with every single city belonging to the group of the most efficient non-US cities. Germany is ranked second by collaborating with 87.50% of the most efficient non-US cities in one of the top 1-5 positions. As compared to that of the United States, the frequency of occurrence of Germany in the top position is only 8.33%. In the case of all articles, the top 1-5 collaborators of the most efficient non-US cities are the United States, Germany, England, France, and Italy. As top collaborators, other countries (like the Netherlands, Australia, Spain, etc.) are rather marginal, primarily appearing in the top 4-5 positions.

In the case of all articles produced in the most efficient US cities (52 out of the top 100 cities), the most frequently occurring countries as collaborators in the top 1-5 positions are Germany, England, China, Canada, and France (Table 4). China, the top collaborator of the most efficient US cities, has surpassed England by almost 2%. The United States has had a traditionally close scientific relationship with Western European countries (especially the United Kingdom) and Canada (Adams 2013), but on the city level, China has recently been occupying a more significant position (Csomós 2018; Tian 2016). Naturally, the top international collaborator of most Chinese cities has been the United States for a long time (He 2009; Wang et al. 2013; Zhang and Guo 1997). If we merge the groups of the most efficient non-US cities and the most efficient US cities into a single group, it turns out that all the co-authors are affiliated with 21 countries occupying one of the top 1-5 positions.

Table 4 illustrates that the international collaboration patterns of the bottom 100 least efficient cities resemble a mixture of the international collaboration patterns of the most efficient non-US cities and US cities. The United States (in the top position in 85% of the cases), Germany, England, France, and China appear in the top 1-5 positions in most cases. However, two facts should be highlighted: 1) As for the international collaboration patterns of the bottom 100 cities, the frequency of occurrence of countries following the United States is much lower than in the case of the most efficient non-US cities. The mean frequency of occurrence of the top 1-5 collaborator countries in articles produced in the most efficient non-US cities is 77.92%. This value is 88.30% in articles produced in the most efficient US cities, but it reaches only 63% in the bottom 100 least efficient cities. 2) The least efficient cities collaborate with a greater number of countries (33) occupying one of the top 1-5 positions than the most efficient cities (21). Many of these countries (e.g., Saudi Arabia, Brazil, Iran, Russia, and South Korea) produce low publishing efficiency; thus, the collaboration has a negative effect on cities' publishing efficiency (i.e., these collaborations result in a smaller number of articles that receive enough citations to belong to the top 1% highly cited articles).

Table 4. Top collaborators* in the case of all articles.

|   | Top collaborators of the most efficient non-US cities occurring in the 1-5 positions | Frequency of occurrence in the 1-5 positions in percentage | Top collaborators of the most efficient US cities occurring in the 1-5 positions | Frequency of occurrence in the 1-5 position in percentage | Top collaborators of the least efficient cities occurring in the 1-5 positions | Frequency of occurrence in the 1-5 positions in percentage |
|---|---|---|---|---|---|---|
| 1 | USA | 100.00 | Germany | 98.11 | USA | 98.00 |
| 2 | Germany | 87.50 | England | 98.11 | Germany | 78.00 |
| 3 | England | 75.00 | China | 94.34 | England | 69.00 |
| 4 | France | 75.00 | Canada | 84.91 | France | 39.00 |
| 5 | Italy | 52.08 | France | 66.04 | China | 31.00 |
| 6 | Netherlands | 27.08 | Australia | 15.09 | Australia | 30.00 |
| 7 | Australia | 18.75 | Italy | 15.09 | Japan | 25.00 |
| 8 | Spain | 18.75 | Japan | 5.66 | Canada | 23.00 |
| 9 | China | 16.67 | Netherlands | 5.66 | Italy | 23.00 |
| 10 | Scotland | 8.33 | South Korea | 5.66 | South Korea | 18.00 |

* In this context, collaborators correspond to countries with which co-authors are affiliated.

It is, however, more important to know which countries (more precisely the co-authors affiliated with that country) are the top collaborators of cities (more precisely the authors affiliated with that city) in highly cited articles. According to my hypothesis, countries as the top 1-5 collaborators of cities in highly cited articles differ from those occupying top positions in the total number of articles. The publishing efficiency of cities is heavily influenced by where the top collaborators are in the case of highly cited articles.

Table 5 shows that the collaboration pattern of the most efficient non-US cities in highly cited articles is almost the same as the collaboration pattern that emerged in the total number of articles; however, the relative weight of Germany, France, and England has increased. In the total number of articles, the mean frequency of occurrence of the top 1-5 collaborators was 77.92, while in highly cited articles, this value has increased to 79.17. In highly cited articles produced in the most efficient US cities, the frequency of occurrence of England is 100%, which means that England occupies one of the top 1-5 positions of every single city (in the top position in 57.69% of the cases). Germany has the same frequency of occurrence in highly cited articles than in the total number of articles, but the frequency of occurrence of Canada and especially that of France has significantly increased. China, the third most frequently occurring country in the total number of articles, has vanished from the group of the top collaborators in highly cited articles. This means that, although the total number of articles in US cities shows intense collaboration with China, the collaboration results in only a small number of highly cited articles. In highly cited articles, the mean frequency of occurrence of the most efficient US cities with the top 1-5 collaborators is 81.92%, which is a bit less than in the total number of articles.

Not surprisingly, in highly cited articles, the bottom 100 least efficient cities have a very intense collaboration with the United States. In 98 cities, the United States occupies one of the top 1-5 positions and is in the top position in 79% of the cases. The frequency of occurrence of countries following the United States is much lower than in the most efficient cities. The mean frequency of occurrence of the top 1-5 collaborators in highly cited papers produced in the least efficient cities is only 63.4%. In the top 1-5 positions, the bottom 100 least efficient cities collaborate with a total of 30 countries, while this value in the top 100 most efficient (non-US and US) cities is 16.

Table 5. Top collaborators* in the case of the highly cited articles.

|   | Top collaborators of the most efficient non-US cities occurring in the 1-5 positions | Frequency of occurrence in the 1-5 positions in percentage | Top collaborators of the most efficient US cities occurring in the 1-5 positions | Frequency of occurrence in the 1-5 position in percentage | Top collaborators of the least efficient cities occurring in the 1-5 positions | Frequency of occurrence in the 1-5 positions in percentage |
|---|---|---|---|---|---|---|
| 1 | USA | 100.00 | England | 100.00 | USA | 98.00 |
| 2 | Germany | 89.58 | Germany | 98.08 | Germany | 72.00 |

| 3 | France | 79.17 | France | 88.46 | England | 70.00 |
| 4 | England | 77.08 | Canada | 86.54 | France | 43.00 |
| 5 | Italy | 50.00 | Australia | 36.54 | Australia | 34.00 |
| 6 | Netherlands | 22.92 | Italy | 34.62 | China | 29.00 |
| 7 | Spain | 18.75 | China | 21.15 | Italy | 29.00 |
| 8 | Switzerland | 16.67 | Spain | 9.62 | Spain | 26.00 |
| 9 | Australia | 14.58 | Netherlands | 9.62 | Canada | 25.00 |
| 10 | Canada | 12.50 | Switzerland | 7.69 | Japan | 13.00 |

\* In this context, collaborators correspond to countries with which co-authors are affiliated.

In the case of the highly cited articles, there are fundamental differences between the international collaboration patterns of the most efficient cities and the least efficient cities. Although both groups of cities have roughly the same top collaborators, the least efficient cities collaborate with a much greater number of countries than the most efficient cities. It seems that this difference significantly influences the publishing efficiency of cities.

In conclusion, if co-authors are primarily from countries of the United States, Germany, England, France, Canada, and Italy, which are leading countries in science, articles will have a greater chance to receive enough citations to belong to the top 1% highly cited articles.

**Productivity of Scientific Disciplines in Cities as a Factor Influencing Publishing Efficiency**
Beside the factors detailed above, cities' publishing efficiency is significantly influenced by the productivity of scientific disciplines. The most productive disciplines vary city to city, and the productivity of different disciplines in terms of highly cited articles differs as well (Bornmann et al. 2011). In each city, the most productive disciplines will be revealed both in the case of all articles and in the case of highly cited articles.

For example, in the period from 2014 to 2016, authors from Ann Arbor, Michigan produced articles in 151 disciplines: 8.16% of the 27,322 articles were published in the discipline of physics, 7.41% in engineering, 6.43% in 'science, technology, and other topics', 4.99% in chemistry, 4.91% in psychology, and so on. The greatest number of highly cited articles was produced in quite different disciplines; 15.11% of the 765 highly cited articles were written in 'science, technology, and other topics', 11.27% in general internal medicine, 9.35% in physics, 9.22% in oncology, 5.89% in astronomy and astrophysics, and so on.

To obtain a better understanding of why the publishing efficiency of the most efficient cities and that of the least efficient cities differ significantly, we need to reveal the characteristics of the most productive discipline in those cities. Table 6 shows that, in the case of the top 100 cities, the most productive discipline occurring in the top 1-5 positions is 'science, technology, and other topics'. In the Web of Science, articles published in multidisciplinary journals (e.g., Nature, Science, Proceedings of the National Academy of Sciences of the United States of America, and PlosONE) are classified into the discipline of 'science, technology, and other topics'. It is well-known that articles published in high-impact multidisciplinary journals become highly cited at a very great proportion. For example, 45.67% of all articles published between 2014 and 2016 in *Nature* and 40.44% of all articles published in the same period in *Science* have received enough citations to belong to the top 1% highly cited articles.

In general, articles published in the top 100 most efficient cities can be classified into two major scientific fields: natural sciences (e.g., physics, chemistry, and engineering) and health sciences (e.g., neurosciences and neurology, oncology, and psychology). Contrary to the top 100 cities, most articles produced in the bottom 100 least efficient cities can be classified into disciplines that are natural sciences, while the field of health sciences is almost absent. In the case of the least efficient cities, oncology, a discipline in health sciences, is the most frequently occurring discipline with a frequency of occurrence of only 12% (i.e., it occurs in the top 1-5 positions in only 12% of the cities). In contrast to health sciences, natural sciences (e.g., chemistry, engineering, physics, and material science) produce a very high frequency of occurrence (Table 6). Chemistry is in the top 1-5 positions in almost every bottom

100 city, and it occupies the top position in 54% of the cases. This means that, in more than half of the least efficient cities, chemistry is the most productive research area.

Table 6. Most productive scientific disciplines in all articles.

|   | The most productive scientific disciplines occurring in the top 1-5 positions in the most efficient cities | Frequency of occurrence in the 1-5 positions in percentage | The most productive scientific disciplines occurring in the top 1-5 positions in the least efficient cities | Frequency of occurrence in the 1-5 positions in percentage |
|---|---|---|---|---|
| 1 | Science, Technology, and Other Topics | 84.00 | Chemistry | 99.00 |
| 2 | Physics | 63.00 | Engineering | 85.00 |
| 3 | Neurosciences and Neurology | 47.00 | Physics | 84.00 |
| 4 | Chemistry | 45.00 | Materials Science | 80.00 |
| 5 | Engineering | 41.00 | Science, Technology, and Other Topics | 54.00 |
| 6 | Astronomy and Astrophysics | 38.00 | Mathematics | 15.00 |
| 7 | Oncology | 29.00 | Environmental Sciences and Ecology | 12.00 |
| 8 | Environmental Sciences and Ecology | 21.00 | Oncology | 12.00 |
| 9 | Psychology | 20.00 | Pharmacology and Pharmacy | 11.00 |
| 10 | Materials Science | 16.00 | Agriculture | 7.00 |

In the case of the highly cited articles published in the top 100 most efficient cities, the discipline of 'science, technology, and other topics' is even more dominant; it is in the top 1-5 positions in 91% of all cities but occurs in the top position in only 20% of the cases. In 35% of the top 100 cities, general internal medicine occupies the top position but ranked second based on the aggregate frequency of occurrence (Table 7). In highly cited articles produced in the most efficient cities, both the number and frequency of occurrence of health disciplines are greater than in all articles. When examining all articles produced in the most efficient cities, general internal medicine is in the top 1-5 positions in only 5% of cases, but in the highly cited articles, this value increases to 69%. Furthermore, the frequency of occurrence of oncology and the discipline of the cardiovascular system and cardiology increased by more than 50%.

In highly cited articles produced in the bottom 100 least efficient cities, most of the dominant disciplines are in natural sciences. In the least efficient cities, the discipline of 'science, technology, and other topics' occupies the top position, but its frequency of occurrence is less than in the most efficient cities.

Table 7. Most productive scientific disciplines in highly cited articles.

|   | The most productive scientific disciplines occurring in the top 1-5 positions in the most efficient cities | Frequency of occurrence in the 1-5 positions in percentage | The most productive scientific disciplines occurring in the top 1-5 positions in the least efficient cities | Frequency of occurrence in the 1-5 positions in percentage |
|---|---|---|---|---|
| 1 | Science, Technology, and Other Topics | 91 | Science, Technology, and Other Topics | 74 |
| 2 | Physics | 69 | Chemistry | 66 |
| 3 | General Internal Medicine | 69 | Physics | 56 |
| 4 | Astronomy and Astrophysics | 54 | Engineering | 54 |
| 5 | Oncology | 42 | General Internal Medicine | 37 |
| 6 | Chemistry | 28 | Materials Science | 37 |
| 7 | Cardiovascular System and Cardiology | 21 | Astronomy and Astrophysics | 25 |
| 8 | Biochemistry and Molecular Biology | 18 | Environmental Sciences and Ecology | 20 |
| 9 | Environmental Sciences and Ecology | 15 | Oncology | 20 |
| 10 | Neurosciences and Neurology | 14 | Mathematics | 17 |

When we examined the international collaboration patterns in both the cases of all articles and in highly cited articles produced in the top 100 cities and produced in the bottom 100 cities, respectively, we found that they differ in the frequency of occurrence of the top collaborators. However, the countries with which they collaborate (i.e., the location of co-authors) were primarily the same. As for the scientific disciplines, there are significant differences between the top 100 cities and the bottom 100 cities in not only the frequency of occurrence of the most productive disciplines but also in the disciplines themselves. In the most efficient cities, highly cited articles are produced in disciplines that are in natural sciences and health sciences to almost the same degree, while, in the least efficient cities, health disciplines are rather marginal. Furthermore, the frequency of occurrence of the discipline of 'science, technology, and other topics' is much higher in articles produced in the most efficient cities than in articles produced in the least efficient cities. This fact suggests that, in the most efficient cities, a greater number of articles are published in high-impact multidisciplinary journals than in the least efficient cities.

**Discussion and Conclusion**

In this paper, I examined whether there were general factors influencing cities' publishing efficiency (i.e., the ratio of highly cited articles to all articles produced in that city). I have found the following five fundamental factors: the dominance of the English language, cities' economic development level, the location of excellent organisations, cities' international collaboration patterns, and the productivity of scientific disciplines.

The dominance of the English language seems to be one of the most (if not the most) significant factors influencing cities' publishing efficiency. About three-quarters of the most efficient cities are in countries in the Anglosphere–Core, and the rest of are in Northern and Western European countries. Contrary to the most efficient cites, 99% of the least efficient cities are in countries outside the Anglosphere–Core.

The economic development level of cities (derived from country-level data) as a factor influencing the publishing efficiency seems less significant than the dominance of the English language. Results show that 98% of the most efficient cities are in high-income countries. It might suggest that there is a correlation between cities' high-income level and cities' high publishing efficiency, but it turned out that one-third of the least efficient cities were also located in high-income countries. The reason for this is that countries that are home to cities with low efficiency but high-income level do not belong to the Anglosphere, reinforcing the fact that the dominance of the English language as a factor has a greater significance in influencing cities' publishing efficiency than the cities' economic development level has.

It is well-known fact that scientific publications are primarily produced by universities. We can assume that the most efficient cities should be home to the most excellent universities in the world, while excellent universities are not expected to be in the least efficient cities. Results show that this hypothesis is basically correct, at least when we focus on the location of excellent universities in the least efficient cities. However, the picture is more complex in the case of the most efficient cities, because half of those cities are not home to excellent universities. Moreover, many excellent universities are in cities that are not the most efficient cities. The reason for this is that there are many towns and small or mid-sized cities that are home to world-renowned national or international research institutions producing even higher publishing efficiency than excellent universities. These settlements are all characterised by the fact that they are within metropolitan areas, while the research institutions they host operate under the umbrella of excellent universities.

In the case of the highly cited articles, an overlap can be detected between the international collaboration patterns of the most efficient cities and the least efficient cities. In both cases, the top collaborators are the United States (primarily in the top position), Germany, England, France, Canada, and Australia/Italy. If we merely focus on who the top collaborators of cities are, we cannot predict whether its publishing efficiency will be high. However, the magnitude of the collaboration intensity between cities (more precisely the authors affiliated with those cities) and the leading countries in science (more precisely the co-authors located in those countries) even more significantly influences cities' publishing efficiency. The higher the collaboration intensity is, the more likely it is that cities will produce high publishing efficiency.

In the most efficient cities, highly cited articles are produced in disciplines of natural sciences and health sciences to the same degree. In the least efficient cities, almost all highly cited articles are produced in the field of natural sciences (primarily in chemistry), while hardly any articles are published in health sciences. In the case of both groups of cities, 'science, technology, and other topics' is the most frequently occurring discipline in highly cited articles; however, its frequency of occurrence in articles produced in the most efficient cities is much higher than in the least efficient cities.

Based on the above research results, we can draw the conclusion that a city's publishing efficiency will be high if meets the following conditions:

1) It is in a country in the Anglosphere–Core;
2) It is in a high-income country;
3) It is home to excellent universities and/or world-renowned research institutions;
4) Researchers affiliated with that city most intensely collaborate with researchers affiliated with cities in the United States, Germany, England, France, Canada, and Australia/Italy; and
5) The most productive scientific disciplines of highly cited articles are 'science, technology, and other topics' (i.e., most articles are published in high-impact multidisciplinary journals), disciplines in health sciences (especially general internal medicine and oncology), and disciplines in natural sciences (especially physics, astronomy, and astrophysics).

Approximately 60% of the top 100 most efficient cities meet the above criteria, but if we expand the geographical dimension beyond the Anglosphere, 86% of the top 100 cities will meet the criteria.

Most of the bottom 100 least efficient cities are in countries outside the Anglosphere. If we do not consider the determinant significance of the linguistic factor, the patterns of the Japanese, South Korean, and European cities resemble the patterns of the most efficient cities. All of them are in high-income countries and have more or less similar international collaboration patterns as that of the most efficient cities. Moreover, most of the highly cited articles are produced in similar disciplines (although disciplines in natural sciences are overrepresented). Naturally, there are several excellent universities and research institutions in Japanese and South Korean cities (especially in Tokyo, Kyoto, Nagoya, Osaka, and Seoul); yet, they produce low publishing efficiency.

The question is: What can the city administration do to increase the city's performance in science (e.g., to increase the city's publishing efficiency)? Naturally, cities have limited opportunities to compete for components of the science establishment. Universities, hospitals and most governmental research institutions are generally tied to their original loci. However, cities can compete to attract innovation-oriented companies, high tech firms, and R&D facilities of multinational companies by for example establishing science parks. The positive effect of this process on the city's performance in science can be observed in the example of Beijing (Andersson et al. 2014; Liefner et al. 2006; Zhou 2005). Furthermore, cities can compete to acquire cutting-edge international research facilities. For example, in 2009, founding member states of the European Spallation Source (ESS) (the most powerful linear proton accelerator in the world) decided to support for placing ESS in Lund, selecting it from the competition of three European cities. The ESS will attract thousands of researchers from all over the world to Lund.

In conclusion, if a city provides an innovative environment that is based upon the collaboration of universities, high tech firms and research facilities and creates an attractive milieu for the creative class, it is predictable that its performance in science will increase.

**Dataset**
Full dataset on cities' publishing efficiency is available at Harvard Dataverse (doi.org/10.7910/DVN/O22H8D).

**Acknowledgements**
The publication is supported by the EFOP-3.6.1-16-2016-00022 project. The project is co-financed by the European Union and the European Social Fund.

Appendix 1. The top 100 most efficient cities

| Rank | Country | City | Efficiency | Rank | Country | City | Efficiency |
|------|---------|------|------------|------|---------|------|------------|
| 1 | France | Villejuif | 6.174 | 51 | USA | Los Angeles, CA | 2.948 |
| 2 | USA | Menlo Park, CA | 5.676 | 52 | USA | Chapel Hill, NC | 2.937 |
| 3 | USA | Princeton, NJ | 4.978 | 53 | Canada | Victoria, BC | 2.924 |
| 4 | USA | Cambridge, MA | 4.670 | 54 | UK | Dundee | 2.916 |
| 5 | USA | Stanford, CA | 4.658 | 55 | USA | Philadelphia, PA | 2.907 |
| 6 | Saudi Arabia | Jeddah | 4.541 | 56 | UK | Leicester | 2.906 |
| 7 | USA | Santa Cruz, CA | 4.430 | 57 | UK | Edinburgh | 2.898 |
| 8 | USA | Pasadena, CA | 4.400 | 58 | USA | Research Triangle Park, NC | 2.897 |
| 9 | USA | San Francisco, CA | 3.993 | 59 | South Africa | Cape Town | 2.896 |
| 10 | USA | Berkeley, CA | 3.932 | 60 | Netherlands | Wageningen | 2.886 |
| 11 | USA | Upton, NY | 3.920 | 61 | Germany | Garching bei München | 2.877 |
| 12 | USA | Bethesda, MD | 3.912 | 62 | USA | Baltimore, MD | 2.866 |
| 13 | USA | Seattle, WA | 3.870 | 63 | Switzerland | Lausanne | 2.866 |
| 14 | USA | Rochester, MN | 3.830 | 64 | Denmark | Copenhagen | 2.855 |
| 15 | USA | Santa Barbara, CA | 3.778 | 65 | USA | Rochester, NY | 2.852 |
| 16 | USA | Boston, MA | 3.742 | 66 | USA | Houston, TX | 2.846 |
| 17 | USA | Greenbelt, MD | 3.679 | 67 | Estonia | Tartu | 2.842 |
| 18 | USA | Rockville, MD | 3.667 | 68 | USA | Providence, RI | 2.840 |
| 19 | USA | Richland, WA | 3.618 | 69 | USA | Denver, CO | 2.837 |
| 20 | Switzerland | Geneva | 3.566 | 70 | USA | Birmingham, AL | 2.826 |
| 21 | USA | New Haven, CT | 3.565 | 71 | South Africa | Durban | 2.818 |
| 22 | UK | Oxford | 3.427 | 72 | France | Clermont-Ferrand | 2.802 |
| 23 | USA | Durham, NC | 3.400 | 73 | USA | Ann Arbor, MI | 2.800 |
| 24 | USA | Evanston, IL | 3.388 | 74 | Italy | Ferrara | 2.797 |
| 25 | UK | Didcot | 3.366 | 75 | USA | Cleveland, OH | 2.788 |
| 26 | USA | Boulder, CO | 3.288 | 76 | Canada | Hamilton, ON | 2.768 |
| 27 | USA | Dallas, TX | 3.272 | 77 | UK | Southampton | 2.758 |
| 28 | USA | New York, NY | 3.262 | 78 | UK | Cardiff | 2.738 |
| 29 | Italy | Perugia | 3.219 | 79 | UK | Exeter | 2.738 |
| 30 | USA | Riverside, CA | 3.201 | 80 | USA | San Diego, CA | 2.734 |
| 31 | Germany | Heidelberg | 3.177 | 81 | USA | Hanover, NH | 2.715 |
| 32 | UK | Cambridge | 3.171 | 82 | Germany | Mainz | 2.714 |
| 33 | UK | Brighton | 3.162 | 83 | USA | Gaithersburg, MD | 2.691 |
| 34 | USA | Nashville, TN | 3.122 | 84 | USA | Worcester, MA | 2.687 |
| 35 | France | Créteil | 3.111 | 85 | Switzerland | Villigen | 2.686 |
| 36 | Israel | Rehovot | 3.096 | 86 | UK | Birmingham | 2.685 |
| 37 | USA | Portland, OR | 3.091 | 87 | Denmark | Lyngby | 2.684 |
| 38 | USA | Palo Alto, CA | 3.080 | 88 | Germany | Bonn | 2.678 |
| 39 | Switzerland | Basel | 3.050 | 89 | Canada | Toronto, ON | 2.660 |
| 40 | Italy | Trieste | 3.036 | 90 | UK | Newcastle | 2.658 |
| 41 | USA | St. Louis, MO | 3.029 | 91 | Switzerland | Bern | 2.657 |
| 42 | Netherlands | Rotterdam | 3.028 | 92 | USA | Amherst, MA | 2.652 |
| 43 | Canada | Vancouver, BC | 3.024 | 93 | USA | Eugene, OR | 2.650 |
| 44 | UK | Norwich | 3.006 | 94 | Netherlands | Amsterdam | 2.643 |
| 45 | USA | Aurora, CO | 3.004 | 95 | USA | Chicago, IL | 2.638 |
| 46 | USA | Atlanta, GA | 2.982 | 96 | Germany | Essen | 2.627 |
| 47 | UK | Lancaster | 2.976 | 97 | Belgium | Brussels | 2.614 |
| 48 | Netherlands | Nijmegen | 2.963 | 98 | Italy | Pavia | 2.611 |
| 49 | USA | San Antonio, TX | 2.961 | 99 | USA | Winston-Salem, NC | 2.594 |
| 50 | France | Gif-sur-Yvette | 2.955 | 100 | USA | Tallahassee, FL | 2.591 |

Appendix 2. The bottom 100 least efficient cities

| Rank | Country | City | Efficiency | Rank | Country | City | Efficiency |
|---|---|---|---|---|---|---|---|
| 1 | Tunisia | Sfax | 0.132 | 51 | Japan | Shizuoka | 0.626 |
| 2 | Russia | Yekaterinburg | 0.161 | 52 | China | Nanning | 0.632 |
| 3 | South Korea | Cheonan | 0.260 | 53 | India | Hyderabad | 0.632 |
| 4 | Iran | Shiraz | 0.268 | 54 | Japan | Saitama | 0.634 |
| 5 | Romania | Iași | 0.273 | 55 | Japan | Kawasaki | 0.649 |
| 6 | India | Kharagpur | 0.283 | 56 | Brazil | Recife | 0.654 |
| 7 | China | Mianyang | 0.325 | 57 | Italy | Messina | 0.661 |
| 8 | Poland | Lublin | 0.333 | 58 | Egypt | Cairo | 0.670 |
| 9 | Brazil | São Carlos | 0.333 | 59 | Turkey | Istanbul | 0.673 |
| 10 | China | Wenzhou | 0.348 | 60 | China | Changzhou | 0.682 |
| 11 | India | Varanasi | 0.399 | 61 | South Korea | Yongin | 0.682 |
| 12 | China | Shijiazhuang | 0.416 | 62 | China | Kunming | 0.689 |
| 13 | South Korea | Cheongju | 0.424 | 63 | Pakistan | Lahore | 0.690 |
| 14 | Japan | Gifu | 0.436 | 64 | Japan | Sapporo | 0.695 |
| 15 | Iran | Tabriz | 0.444 | 65 | Argentina | Córdoba | 0.720 |
| 16 | Chile | Concepción | 0.452 | 66 | Japan | Kanazawa | 0.732 |
| 17 | Brazil | Curitiba | 0.454 | 67 | Poland | Gdańsk | 0.736 |
| 18 | Japan | Kumamoto | 0.456 | 68 | Poland | Wrocław | 0.736 |
| 19 | Malaysia | Serdang | 0.462 | 69 | China | Qingdao | 0.737 |
| 20 | Tunisia | Tunis | 0.484 | 70 | Ukraine | Kiev | 0.749 |
| 21 | Egypt | Giza | 0.487 | 71 | China | Jinan | 0.749 |
| 22 | China | Nantong | 0.494 | 72 | China | Xinxiang | 0.754 |
| 23 | Israel | Beer-Sheva | 0.501 | 73 | India | New Delhi | 0.755 |
| 24 | Japan | Ibaraki | 0.503 | 74 | Poland | Łódź | 0.756 |
| 25 | India | Kanpur | 0.513 | 75 | China | Ningbo | 0.758 |
| 26 | China | Baoding | 0.516 | 76 | India | Bangalore | 0.783 |
| 27 | Turkey | Konya | 0.535 | 77 | South Korea | Jinju | 0.783 |
| 28 | South Korea | Busan | 0.537 | 78 | Turkey | Ankara | 0.791 |
| 29 | Iran | Tehran | 0.550 | 79 | Japan | Chiba | 0.796 |
| 30 | China | Shenyang | 0.551 | 80 | Japan | Sagamihara | 0.798 |
| 31 | Egypt | Alexandria | 0.552 | 81 | South Africa | Pretoria | 0.801 |
| 32 | Japan | Niigata | 0.556 | 82 | Russia | Novosibirsk | 0.804 |
| 33 | France | Villeneuve-d'Ascq | 0.562 | 83 | South Korea | Goyang | 0.804 |
| 34 | Spain | Alicante | 0.563 | 84 | South Korea | Daegu | 0.806 |
| 35 | South Korea | Gwangju | 0.563 | 85 | South Korea | Seoul | 0.807 |
| 36 | South Korea | Jeonju | 0.566 | 86 | China | Nanchang | 0.809 |
| 37 | Brazil | Fortaleza | 0.567 | 87 | China | Taiyuan | 0.810 |
| 38 | Poland | Poznań | 0.568 | 88 | China | Guiyang | 0.813 |
| 39 | Brazil | Viçosa | 0.576 | 89 | India | Roorkee | 0.829 |
| 40 | Turkey | Izmir | 0.587 | 90 | Russia | Moscow | 0.836 |
| 41 | India | Lucknow | 0.587 | 91 | China | Wuxi | 0.840 |
| 42 | Portugal | Aveiro | 0.587 | 92 | Brazil | Porto Alegre | 0.844 |
| 43 | China | Zhengzhou | 0.588 | 93 | Brazil | Florianópolis | 0.855 |
| 44 | China | Guilin | 0.594 | 94 | Russia | Saint Petersburg | 0.856 |
| 45 | China | Yantai | 0.595 | 95 | India | Mumbai | 0.865 |
| 46 | South Korea | Daejeon | 0.596 | 96 | Japan | Sendai | 0.865 |
| 47 | Brazil | Belo Horizonte | 0.601 | 97 | UK | Loughborough | 0.868 |
| 48 | India | Kolkata | 0.606 | 98 | China | Xuzhou | 0.869 |
| 49 | China | Ürümqi | 0.613 | 99 | Brazil | Campinas | 0.884 |
| 50 | India | Chennai | 0.617 | 100 | Poland | Kraków | 0.891 |